\documentstyle[12pt,epsf]{article}

\newcommand{\la}[1]{\label{#1}}

\newlength{\numlen}

\newlength{\indexlength}

\newlength{\llentmp}
\newcommand{\llen}[1]{\settowidth{\llentmp}{$#1$}\hspace{\llentmp}}
 
\newcommand{\be}{\begin{equation}}
\newcommand{\ee}{\end{equation}}
\newcommand{\ba}{\begin{eqnarray}}
\newcommand{\ea}{\end{eqnarray}}

\newcommand{\ie}{{\em i.\,e.\ }}

\newcommand{\etal}{{et al.\ }}
\newcommand{\eq}{Eq.~}
\newcommand{\eqs}{Eqs.~}
\newcommand{\fig}{Fig.~}
\newcommand{\figs}{Figs.~}

\newcommand{\cO}{{\cal O}}
\newcommand{\nr}[1]{(\ref{#1})}
\newcommand{\h}{{\hspace{0.5 cm}}}

\newcommand{\re}{{\rm Re\,}}
\newcommand{\tr}{{\rm Tr\,}}

\newcommand{\fr}[2]{{\frac{#1}{#2}}}
\def\lsim{\raise0.3ex\hbox{$<$\kern-0.75em\raise-1.1ex\hbox{$\sim$}}}
\def\gsim{\raise0.3ex\hbox{$>$\kern-0.75em\raise-1.1ex\hbox{$\sim$}}}

\newcommand{\betapl}{\beta_{\rm pl}}
\newcommand{\betart}{\beta_{\rm rt}}
\newcommand{\betapg}{\beta_{\rm pg}}
\newcommand{\cpl}{c_{\rm pl}}
\newcommand{\crt}{c_{\rm rt}}
\newcommand{\cpg}{c_{\rm pg}}

\begin{document}
\begin{titlepage}
\baselineskip=14pt
\rightline{AZPH-TH/97-13}
\rightline{FSU-SCRI-97-130}
\rightline{IUHET-374}
\rightline{NORDITA-97/84P}

\baselineskip=17pt
\vskip 1.0cm

\centerline{\Large \bf Quenched hadron spectroscopy with }
\centerline{\Large \bf improved staggered quark action}
\bigskip

\centerline{\bf Claude~Bernard }
\centerline{\it
Department of Physics, Washington University, St.~Louis, MO 63130, USA
}
\centerline{\bf Tom~Blum} 
\centerline{\it
Department of Physics, Brookhaven National Lab, Upton, NY 11973, USA
}
\centerline{\bf Thomas~A.~DeGrand}
\centerline{\it
Physics Department, University of Colorado, Boulder, CO 80309, USA
}
\centerline{\bf Carleton~DeTar, Craig~McNeile }
\centerline{\it
Physics Department, University of Utah, Salt Lake City, UT 84112, USA
}
\centerline{\bf Steven~Gottlieb}
\centerline{\it
Department of Physics, Indiana University, Bloomington, IN 47405, USA
}
\centerline{\bf Urs~M.~Heller }
\centerline{\it
SCRI, Florida State University, Tallahassee, FL 32306-4130, USA
}
\centerline{\bf James Hetrick }
\centerline{\it
Physics Department, University of the Pacific, Stockton, CA
95211-0197, USA
}
\centerline{\bf K.~Rummukainen}
\centerline{\it
NORDITA, Blegdamsvej 17, DK-2100 Copenhagen \O, Denmark
}
\centerline{\bf Bob~Sugar }
\centerline{\it
Department of Physics, University of California, Santa Barbara, CA 93106, USA
}
\centerline{\bf Doug~Toussaint }
\centerline{\it
Department of Physics, University of Arizona, Tucson, AZ 85721, USA
}

\bigskip\noindent
We investigate light hadron spectroscopy with an
improved quenched staggered quark action.  We compare the results
obtained with an improved gauge plus an improved quark action, an improved
gauge plus standard quark action, and the standard gauge plus standard quark
action.  Most of the improvement in the spectroscopy results is
due to the improved gauge sector.  However, the
improved quark action substantially reduces violations of
Lorentz invariance, as evidenced by the meson dispersion relations.
\end{titlepage}

\baselineskip=19pt

\section{Introduction}

The precision of numerical lattice QCD simulations with the standard
lattice actions is constrained by the available computational
resources.  In order to keep the duration of the calculation within
manageable bounds, one is forced to use lattice spacings $a$ which may
be too large to accurately describe the continuum physics.  This
problem has been addressed by the development of improved
\cite{Symanzik83,LuscherWeisz,LepageMackenzie} and fixed-point
\cite{DeGrand,perfectKS} actions.  The promise of these actions is to yield good
approximations to the continuum physics with relatively coarse lattice
spacings. (For a recent review, see \cite{Niedermayer96}.)

In the Symanzik improvement scheme \cite{Symanzik83,LuscherWeisz} the
lattice action and fields are improved in powers of the lattice
spacing $a$.  This is achieved by introducing higher dimensional terms
into the action.  In the continuum limit, these terms are irrelevant, but
at a finite cutoff the coefficients of these terms can be tuned so
that the discretization errors of spectral quantities are diminished.
The most straightforward method to determine the coefficients is to
expand the action in a Taylor series in $a$, and cancel the leading
scaling violating terms order by order (tree-level improvement).  
This can be refined by using perturbative analysis or 
non-perturbative numerical methods to determine the coefficients.

In this paper, we study the improvement of the staggered
(Kogut-Susskind) quark lattice QCD action with quenched spectroscopy
calculations.  The improvement is implemented by adding a
third-nea\-rest-neigh\-bor term, first proposed by Naik more than a
decade ago \cite{Naik}.  Some of the preliminary results of this study
have already been published in \cite{milc_naik,fatlinks}.  The same
improvement scheme has been applied to nonzero temperature calculations
by Karsch \etal \cite{Karsch96}.  The gauge configurations used in
this study are generated with an $\cO(a^2)$ one-loop and 
tadpole-improved gauge
action \cite{LuscherWeisz,LepageMackenzie}.  Since our main goal is to
investigate the effects of the fermionic improvement, we compare the
hadronic spectrum obtained with both the unimproved and improved
fermion actions, using the same gauge configurations.  
An excellent baseline for the evaluation of the improvement is provided
by our extensive standard (non-improved)
quenched Kogut-Susskind hadron spectroscopy calculation 
\cite{milcspectrum}.

As opposed to Wilson fermions, the improvement of
the staggered action has attracted relatively little attention.  This
is partly due to the formal complexity of the staggered formulation,
partly to the fact that the standard Wilson fermions have an error
$\cO(a)$, whereas the staggered action is already accurate to this
order.  Nevertheless, the improvement of the staggered action is
highly desirable: the staggered action has a U(1)$\times$U(1) chiral
symmetry, remnant of the full continuum U(4)$\times$U(4) symmetry (for
4 quark flavors).  This symmetry is restored in the continuum limit;
however, for practical values of the lattice spacing a substantial
flavor symmetry breaking remains.  This is a lattice artifact, and it
remains a major problem when one studies the restoration of the
spontaneously broken chiral symmetry at finite temperature.  Moreover,
the very successful $\cO(a^2)$ improvement of the pure gauge action 
makes it very natural to try to bring the quark action to the same accuracy.

This paper is organized as follows: in section 2 we discuss the
improvement of both the gauge and the fermion actions and the
properties of the free fermion actions.  In section 3 we present the
results of the simulations and the comparison of the different
actions.  In particular, we study (a) the $m_N/m_\rho$ mass ratios at
several fixed values of $m_\pi/m_\rho$ as functions of the lattice
spacing, (b) the Lorentz invariance of $\pi$ and $\rho$ meson states,
and (c) the restoration of the flavor symmetry (as determined by the
mass difference of the pseudo-Goldstone and non-Goldstone $\pi$
mesons).  Our conclusions are presented in section 4.

\section{Improvement of the action}

\subsection{The gauge action}

We generate gauge configurations with the tadpole-improved SU(3)
gauge action
\cite{LuscherWeisz,LepageMackenzie,Alford95}:
\be
  S_G = \betapl \sum_{x;\mu<\nu} (1 -  P_{\mu\nu})
      + \betart \sum_{x;\mu\neq\nu} (1 - R_{\mu\nu})
      + \betapg \sum_{x;\mu<\nu<\sigma} (1 -  C_{\mu\nu\sigma})
 \la{gaugeaction}
\ee
where $P$ is the standard plaquette in the $\mu,\nu$ -plane, and $R$
and $C$ denote the real part of the trace of the ordered product of
SU(3) link matrices along $1\times 2$ rectangles and $1\times 1 \times
1$ paths, respectively:
\newlength{\latlength} 
\setlength{\latlength}{1mm} 
\setlength{\unitlength}{\latlength}
\ba
   P_{\mu\nu} &=& \fr13 \re\tr 
	\raisebox{-4\latlength}{\begin{picture}(10,12)
			\put( 0, 0){\vector( 1, 0){7}}
			\put( 0, 0){\line( 1, 0){10}}
			\put(10, 0){\vector( 0, 1){7}}
			\put(10, 0){\line( 0, 1){10}}
			\put(10,10){\vector(-1, 0){7}}
			\put(10,10){\line(-1, 0){10}}
			\put( 0,10){\vector( 0,-1){7}}
			\put( 0,10){\line( 0,-1){10}}
		\end{picture}}  \\
   R_{\mu\nu} &=& \fr13 \re\tr
	\raisebox{-4\latlength}{\begin{picture}(20,12)
			\put( 0, 0){\vector( 1, 0){7}}
			\put( 0, 0){\line( 1, 0){10}}
			\put(10, 0){\vector( 1, 0){7}}
			\put(10, 0){\line( 1, 0){10}}
			\put(20, 0){\vector( 0, 1){7}}
			\put(20, 0){\line( 0, 1){10}}
			\put(20,10){\vector(-1, 0){7}}
			\put(20,10){\line(-1, 0){10}}
			\put(10,10){\vector(-1, 0){7}}
			\put(10,10){\line(-1, 0){10}}
			\put( 0,10){\vector( 0,-1){7}}
			\put( 0,10){\line( 0,-1){10}}
			\multiput(10, 1)(0,1){9}{\circle*{0.1}}
		\end{picture}}  \\
   C_{\mu\nu\sigma} &=& \fr13 \re\tr
	\raisebox{-4\latlength}{\begin{picture}(16,16)
			\put( 0, 0){\vector( 1, 0){7}}
			\put( 0, 0){\line( 1, 0){10}}
			\put(10, 0){\vector( 2, 1){4}}
			\put(10, 0){\line( 2, 1){6}}
			\put(16, 3){\vector( 0, 1){7}}
			\put(16, 3){\line( 0, 1){10}}
			\put(16,13){\vector(-1, 0){7}}
			\put(16,13){\line(-1, 0){10}}
			\put( 6,13){\vector(-2,-1){4}}
			\put( 6,13){\line(-2,-1){6}}
			\put( 0,10){\vector( 0,-1){7}}
			\put( 0,10){\line( 0,-1){10}}
			\multiput(1,0.5)(1,0.5){6}{\circle*{0.1}}
			\multiput(6,4)(0,1){9}{\circle*{0.1}}
			\multiput(7,3)(1,0){9}{\circle*{0.1}}
		\end{picture}} 
\ea
In general, the improvement conditions do not uniquely specify the
form of the action.  For example, at tree-level, adding either
the planar 6-link term or one of several 8-link terms to the standard
action would cancel the $\cO(a^2)$ errors.  However, when the quantum
corrections are calculated with the lattice perturbation theory, then
at least two terms are required to cancel $\cO(g^{2n}a^2)$
errors~\cite{LuscherWeisz}.  The terms in \eq\nr{gaugeaction} provide
the most compact form of the action.

Due to the UV divergence of the tadpole-type graphs in lattice
perturbation theory, operators formally of order $a^n$ in the
expansion of the action are changed to order $a^{n-2m}g^{2m}$ by
quantum effects, depending on the number of tadpole graph
contributions to that particular term.  (In the tadpole contributions
$n\ge 2m$, \ie tadpoles do not introduce additional UV-divergencies.)
The contribution of the tadpole diagrams can be partially taken into account
by absorbing them in the lattice coupling constants.
This is commonly achieved by the definition of the `average
gauge link' from the plaquette, $u_0 \equiv \langle P \rangle^{1/4}$, 
which is strongly dominated by tadpoles, and by replacing $U_{i}(x)\rightarrow
U_{i}(x)/u_0$ in every lattice operator
\cite{LepageMackenzie,Alford95}.  This corresponds to a redefinition of
the lattice gauge coupling $g^2\rightarrow g^2/u_0^4$.

With these ingredients, the coefficients of the action \nr{gaugeaction}
are related by the 1-loop expressions
\cite{Alford95}
\ba
  \betart &=& -\fr{\betapl}{20\,u_0^2} 
	(1+0.4805\alpha_s)  \la{betart} \\ 
  \betapg &=& -\fr{\betapl}{u_0^2} 0.03325 \alpha_s \la{betapg}
\ea
where the strong coupling constant is determined through the 1-loop
relation
\be
  \alpha_s = -4\log(u_0)/3.0684 \, .  \la{alphas}
\ee
The leading errors of this action are of order $\cO(a^2\alpha_s^2,a^4)$.

\subsection{Tree-level improvement of the quark action}

In this work, we study the following fermion action:
\ba
  S_N &=& a^4\sum_{x;\mu} \eta_{\mu}(x)\bar\chi(x) \fr{1}{2a}\bigg\{ 
	   c_1\,\Big[U_{\mu}(x) \chi(x+\mu)
	 - U^\dagger_\mu (x-\mu) \chi(x-\mu)\Big] \nonumber \\ 
      &+& c_2 \,\Big[U_{\mu}(x)U_\mu(x+\mu) U_\mu(x+2\mu)
		 \chi(x+3\mu)\nonumber \\
      & &  \llen{c_2 \Big[}
	- U^\dagger_\mu(x-\mu)U^\dagger_\mu(x-2\mu)U^\dagger_\mu(x-3\mu)
	 \chi(x-3\mu)\Big] \bigg\}   \la{naik} \\
      &+& a^4 m_q \sum_x \bar\chi(x)\chi(x)\,, \nonumber
\ea
where the phase factor $\eta_\mu(x) = (-1)^{(x_0 + x_1 \ldots
x_{\mu-1})}$.  The standard Kogut-Susskind (staggered) action is
obtained with coefficients $c_1 = 1$ and $c_2 = 0$.  At tree
level, the action is $\cO(a^2)$ accurate when $c_1 = 9/8$ and $c_2 =
-1/24$.  In this case, the {\em difference} from the Kogut-Susskind
action is a discrete version of the 3rd order derivative:
\be
  \fr{1}{8}\fr{f(x+\hat\mu) - f(x-\hat\mu)}{2a} - 
  \fr{1}{24}\fr{f(x+3\hat\mu) - f(x-3\hat\mu)}{2a} = 
- \fr{a^2}{6}\,\partial_\mu^3f(x) +\cO(a^4)\,.
\ee
The staggered action with a third nearest neighbor term was originally
proposed by Naik \cite{Naik}.  However, he was studying the
improvement of the Dirac-K\"ahler action, which has a different
coupling to the gauge fields than the action in \eq\nr{naik}.
The Dirac-K\"ahler action lacks the exact $U(1)\times U(1)$ -symmetry
enjoyed by the action \nr{naik}, and the bare quark mass has to
be additively renormalized.  These properties make use of the
Dirac-K\"ahler action much less appealing than the Kogut-Susskind
action.  Nevertheless, in the following we shall call the action
\nr{naik} the Naik action.

The (one-component) Grassmann field $\chi$ describes 4 flavors of
Dirac fermions in the continuum limit.  This is not transparent in
\eq\nr{naik}, nor can one easily identify the leading irrelevant terms
when the continuum limit is taken.  At the free fermion level, perhaps
the easiest way to see this is to use the following transformation
\cite{KSmomentum}: in momentum space, we decompose the
momentum vector  $k = p + \pi A/a$, $A_\mu = 0$ or 1, and we restrict 
$0 \le k <\pi/a$.  A new
(16-component) fermion field $\psi$ is defined as
\ba
  \psi(p) &=& 
	\fr18 \sum_{A,B} (-1)^{A\cdot B} \Gamma_A \chi(p+\pi B/a) \\
  \bar\psi(p) &=& 
	\fr18 \sum_{A,B} (-1)^{A\cdot B} \Gamma^\dagger_A 
	\bar\chi(p+\pi B/a)\,.
\ea
where
\be
  \Gamma_A = \gamma_0^{A_0} \gamma_1^{A_1} \gamma_2^{A_2} \gamma_3^{A_3}\,.
\ee
In terms of field $\psi$, the free action \nr{naik} becomes 
\be
  S_{\rm free} = 
        \sum_p \bar\psi(p) \left[
	\sum_\mu \gamma_\mu \fr{i}{a} \left( c_1 \sin p_\mu a
		+  c_2 \sin 3p_\mu a \right)
	+ m\right] \psi(p)\,. \la{freeaction}
\ee
This form of the action is flavor diagonal; however, if we perform an
inverse Fourier transform, the derivative term becomes nonlocal.  The
Kogut-Susskind action ($c_2=0$) has leading $\cO(a^2)$ errors.
The coefficients $c_1 = 9/8$ and $c_2 = -1/24$ for the Naik action
are readily recovered from
\eq\nr{freeaction} by expanding the trigonometric functions.

When the gauge fields are included one cannot transform the action
\nr{naik} to the form in \eq\nr{freeaction}.  It is not at all obvious
that the interacting Kogut-Susskind action is still $\cO(a)$
-accurate.  In order to see the flavor structure more clearly, one
usually performs the (local) transformation originally proposed by
Kluberg-Stern \etal \cite{KlubergStern}.  It transforms the 1-component
staggered field $\chi$ to a hypercubic 16-component `quark field' (4
flavors of 4-component Dirac spinors), which lives on a lattice with
twice the original lattice spacing.  The quark field action cannot
be written in a compact form, but when expanded in powers of the
lattice spacing $a$ it has apparent dimension-5 terms 
(giving rise to $\cO(a)$ errors).

However, the Kogut-Susskind action does {\em not\,} have on-shell
$\cO(a)$ errors.  This has been shown by Sharpe \cite{Sharpe} and Luo
\cite{Luo96,Luo97} by a generalization of the Kluberg-Stern \etal
transformation.  The leading scaling violations start at $\cO(a^2)$.
In order to cancel them, one has to add dimension-6 terms to the
action; these terms have been classified by Luo \cite{Luo97}.  The terms
fall into two classes: $\bar\psi D^3 \psi$ terms, where $D^3$ is a
generic combination of 3 covariant derivatives (and to which
class the `Naik term' in \eq\nr{naik} belongs), and 4-fermion terms.
Unfortunately, even in the simplest form, the action has 15 dimension-6
terms with --- so far --- unknown coefficients.  Therefore, we limit ourselves
here to a much more modest goal and study the degree of improvement
possible to obtain with the action \nr{naik}, bearing in mind
that this action cannot cancel all of the $\cO(a^2)$ errors, but
only the ones present already for free fermions.

As with the gauge action, we may improve the action \nr{naik} beyond the
tree-level by taking into account the modifications due to gluon tadpoles:
with the replacement $U\rightarrow U/u_0$, the coefficients $c_i$ in action
\nr{naik} become
\be
  c_1 = \fr{9}{8u_0} \h\h  c_2 = -\fr{1}{24u_0^3} \,.  \la{coeffs}
\ee
In this work we use the quark action defined by
\eqs(\ref{naik},\ref{coeffs}).  In the nonzero temperature calculation
in Ref.\ \cite{Karsch96} the action \nr{naik} was used with the
`tree-level' values.

\subsection{Properties of the free quark action}\label{sec:properties}

The free quark dispersion relation $E({\bf p})$ can be found from
\eq\nr{freeaction}, by solving for the poles of the Euclidean
propagator and using the identification $E = \re ip_0$.  In
\fig\ref{figdisp} we show the massless quark dispersion relations for
the standard Kogut-Susskind and Naik actions.
For comparison, we also show the Wilson fermion action dispersion
relation.  The Naik action follows the continuum dispersion relation
$E = |{\bf p}|$ much better than the standard Kogut-Susskind action up
to $|{\bf p}| \sim 1.8/a$, not to mention the Wilson action (with
the Wilson parameter $r=1$).  Note that
for massless free quarks both the Wilson and the Kogut-Susskind
actions have $\cO(a^2)$ leading errors.  Due to the third nearest
neighbor coupling in the imaginary time direction, unphysical {\em
ghost\,} branches (with complex $ip_0$) appear in the 
dispersion relation.  These states
will become infinitely massive when $a\rightarrow 0$.
\begin{figure}[tb]
\vspace*{-1.5cm}
\epsfysize=15cm
\centerline{\epsffile{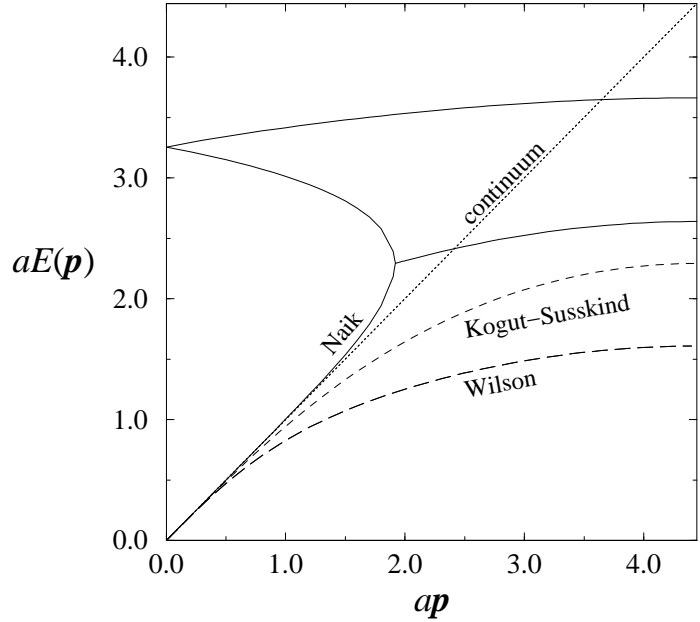}}
\vspace*{-5.6cm}
\caption[a]{The dispersion relation $E(\bf p)$ for massless free
quarks with different fermion actions.  The momentum $\bf p$ is to the
spatial direction $(1,1,0)$, and the dispersion relations are plotted
up to the end of the Brillouin zone.}  \la{figdisp}
\end{figure}

\begin{figure}[tb]
\vspace*{-0.5cm}
\centerline{
\epsfysize=13cm\epsffile{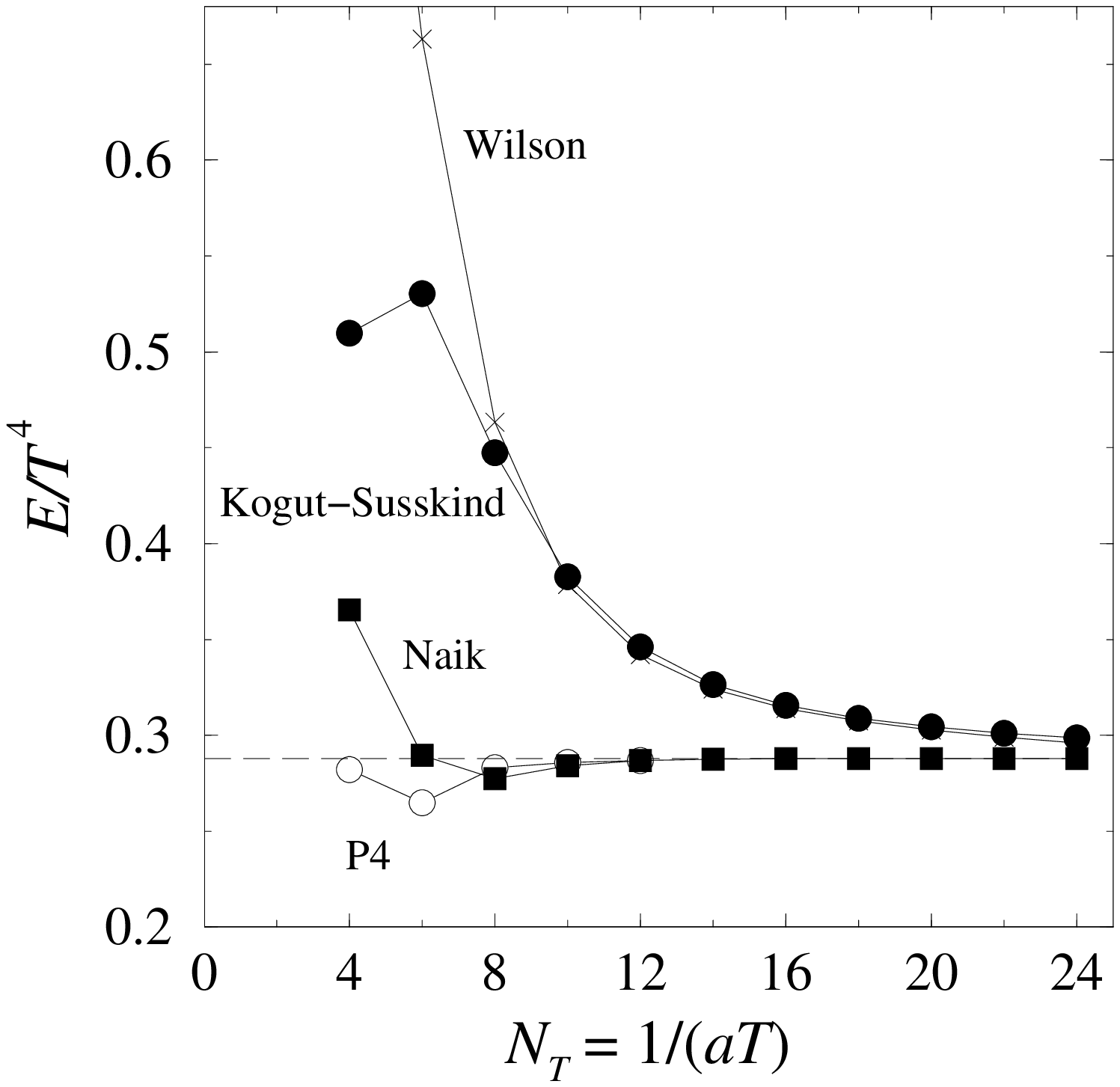}%
\hspace*{-2cm}\epsfysize=13cm\epsffile{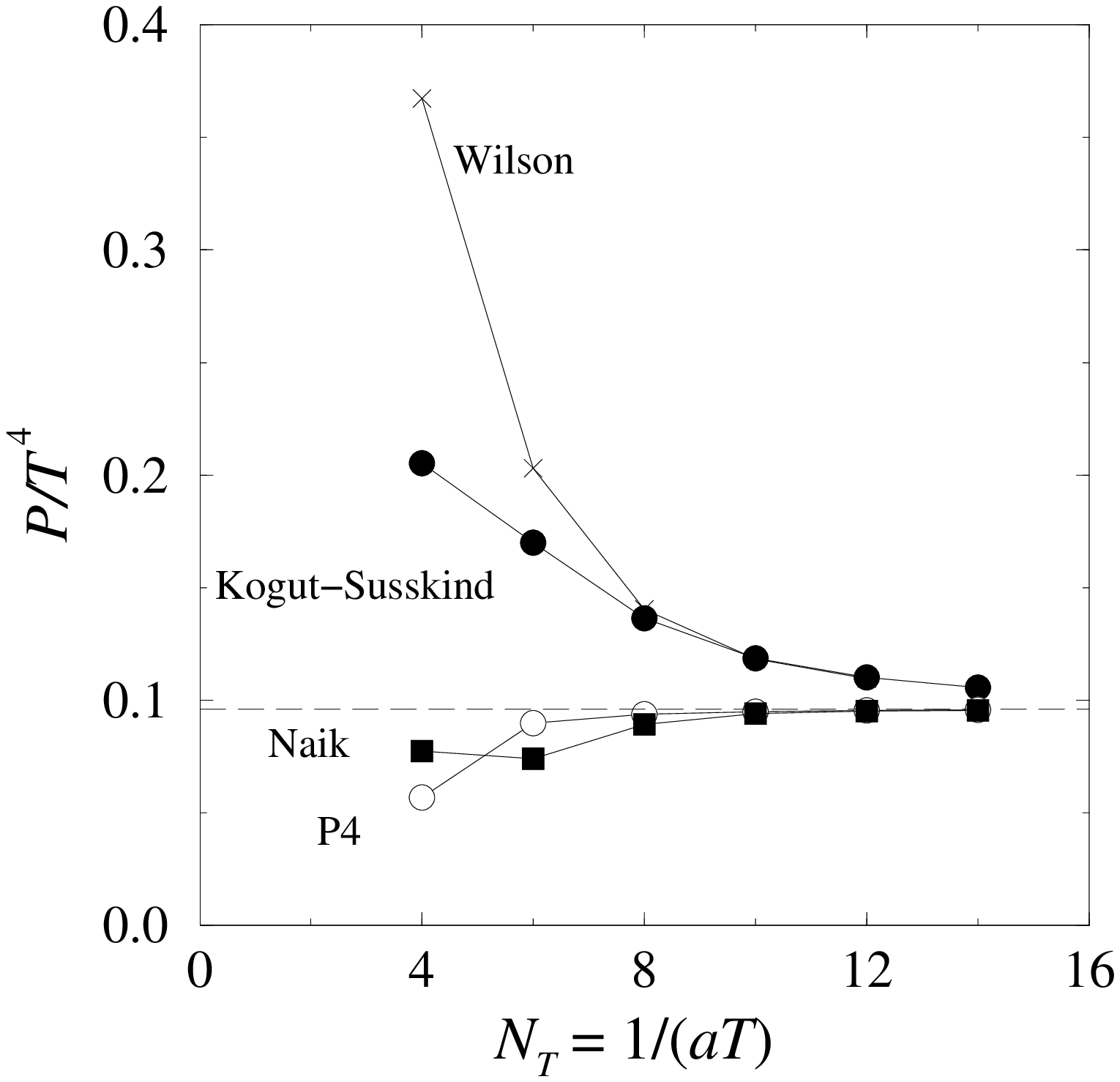}
}
\vspace*{-4cm}
\caption[a]{The energy (left) and the pressure (right) per fermion
degree of freedom for free Kogut-Susskind, Naik, Wilson and ``P4''
\cite{p4} fermions as a function of $N_T = 1/(aT)$.  The continuum
values are shown with dashed lines.} \la{figenergy}
\end{figure}

For free fermions, the thermal energy can be easily calculated from $E
= T^2 \partial \log Z/\partial T $.  In the imaginary time formalism, the
temperature $T = 1/(N_Ta)$, the inverse temporal extent of the lattice.
Under the assumption that the free energy is proportional to the
volume, the pressure is $P = T \partial\log Z/\partial V$.  In
\fig\ref{figenergy} we show $E/T^4$ and $P/T^4$ for free
Kogut-Susskind, Wilson and Naik fermions as functions of the inverse
lattice spacing.  Also shown are the results from the Bielefeld ``P4''
staggered action \cite{p4}: like the Naik action, it contains a 3rd
nearest neighbor coupling, but in this case the neighbors are coupled
along L-shaped paths (the Naik
$\bar\psi(x)U_\mu(x)U_\mu(x+\hat\mu)U_\mu(x+2\hat\mu)\psi(x+3\hat\mu)$
-terms are replaced with terms of form
$\bar\psi(x)U_\mu(x)U_\nu(x+\hat\mu)U_\nu(x+\hat\mu+\hat\nu)
\psi(x+\hat\mu+2\hat\nu)$, with $\nu\ne\mu$).  The P4 action yields
the same tree-level improvement as the Naik action.  (For a comparison
with a renormalization group improved free staggered quark action, see the
last paper of Ref.~\cite{perfectKS}.)

The energy and pressure of the Naik fermions approach the continuum
ideal fermion gas limits much faster than the standard Kogut-Susskind
action.  Indeed, the Bielefeld group \cite{Karsch96} reported an
improved thermodynamic behavior even when the interacting gauge
fields are included in a dynamical quark Monte Carlo simulation.

\section{The simulations and the results}

\subsection{Hadron spectrum}

The parameters of the action \nr{gaugeaction} used in the generation
of the quenched configurations are shown in Table~\ref{tab:runs}.

\begin{table}[ht]
\centerline{
\begin{tabular}{llll}
\hline
 $\betapl$ & $u_0$ & volume & $N_{\rm conf.}$ \\
\hline
 6.8  & 0.8261 & $16^3\times 32$ & 199 \\
 7.1  & 0.8441 & $14^3\times 28$ & 203 \\
 7.4  & 0.8629 & $16^3\times 32$ & 200 \\
 7.6  & 0.8736 & $16^3\times 32$ & 100 \\
 7.75 & 0.8800 & $16^3\times 32$ & 200 \\
 7.9  & 0.8848 & $16^3\times 32$ & 200 \\
\hline
\end{tabular}}
\caption[0]{The parameters of the runs.  $\betart$ and $\betapg$ can
be obtained through \eqs(\ref{betart}--\ref{alphas}).}\la{tab:runs}
\end{table}

We measure the masses of the nucleon, the Goldstone pion $\pi$,
(corresponding to the (spontaneously broken) explicit U(1) chiral
symmetry of the action \nr{naik}), the non-Goldstone (``SC'')
pion $\pi_2$, and the $\rho$ and $\rho_2$ mesons.  The masses were
calculated both with the Naik and the Kogut-Susskind
actions.  For each lattice and propagator, we use four wall source planes.
In each case, the hadron propagators were measured with 5--6 bare
quark masses $am_q = 0.005$ -- 0.32\@; the hadron masses are shown in
Tables \ref{table:mass1}--\ref{table:mass3}.

Throughout the analysis we quantify the performance of the improved
actions by comparing the results against a non-improved benchmark --- an
extensive standard quenched Kogut-Susskind hadron spectroscopy study by the
MILC collaboration \cite{milcspectrum}.  In particular, we use
$\beta_{\rm Wilson} = 6/g^2 = 5.54$ ($16^3$), 5.7 ($24^3$), 5.85
($24^3$) and 6.15 ($32^3$) lattices (with the spatial volume in
parentheses).

\begin{table}[t]
\centerline{
\begin{tabular}{|l|lllll|}
\multicolumn{6}{l}{Naik, $\betapl=6.8$, $16^3\times 32$} \\
\hline
$am_q$ &$\pi$    & $\pi_2$ & $\rho$    & $\rho_2$ & Nucleon   \\
\hline
0.02 &0.343438(70) & 1.4(2)  & 1.537(15)  & 2.06(18)  & 2.301(50)   \\
0.04 &0.484458(64) & 1.53(3) & 1.617(24)  & 2.021(72) & 2.387(19) \\
0.08 &0.682699(82) & 1.86(6) & 1.6964(91) & 2.098(29) & 2.730(38)   \\
0.16 &0.962771(60) & 2.12(3) & 1.8581(75) & -         & 3.032(49)   \\
0.32 &1.365621(51) & -       & 2.061(27)  & -         & 3.589(14) \\
\hline
\multicolumn{6}{l}{Kogut-Susskind, $\betapl=6.8$, $16^3\times 32$} \\
\hline
$am_q$ &$\pi$    & $\pi_2$ & $\rho$    & $\rho_2$ & Nucleon   \\
\hline
0.02 &0.350547(81)& 1.411(75)& 1.405(13)  & 1.73(13)  &2.1632(93) \\
0.04 &0.492871(80)& 1.474(35)& 1.4308(61) & 1.642(48) &2.134(64)  \\
0.08 &0.689935(67)& 1.524(16)& 1.4816(25) & 1.660(17) &2.3286(59) \\
0.16 &0.959548(63)& 1.640(10)& 1.5772(37) & 1.7389(56)&2.447(33)  \\
0.32 &1.325654(74)& 1.8896(43)&1.75802(52)& 1.9078(25)&2.7434(79) \\
\hline
\multicolumn{6}{l}{} \\
\multicolumn{6}{l}{Naik, $\betapl=7.1$, $14^3\times 28$} \\
\hline
$am_q$ &$\pi$     & $\pi_2$  & $\rho$    & $\rho_2$ & Nucleon   \\
\hline
0.02 & 0.35637(12) & 1.289(39) & 1.573(44) & 1.595(67) & 2.16(13)    \\
0.04 & 0.50074(20) & 1.473(59) & 1.568(18) & 1.651(38) & 2.342(51)   \\
0.08 & 0.70269(17) & 1.585(26) & 1.6188(70) & 1.822(23) & 2.459(31)  \\
0.16 & 0.985219(93) & 1.968(65) & 1.822(12) & 2.0087(83) & 2.759(19) \\
0.32 & 1.389359(81) & 2.874(80) & 2.1188(39) & 2.72(16) & 3.384(74)  \\
\hline
\multicolumn{6}{l}{Kogut-Susskind, $\betapl=7.1$, $14^3\times 28$} \\
\hline
$am_q$ &$\pi$     & $\pi_2$  & $\rho$    & $\rho_2$ & Nucleon   \\
\hline
0.02 & 0.36368(13) & 1.411(77) & 1.338(14) & 1.376(31) & 1.990(37)     \\
0.04 & 0.50958(13) & 1.356(32) & 1.403(18) & 1.443(16) & 2.190(59)     \\
0.08 & 0.70934(13) & 1.361(25) & 1.4481(69) & 1.5301(72) & 2.278(18)   \\
0.16 & 0.97939(11) & 1.592(17) & 1.5680(76) & 1.6681(53) & 2.414(11)   \\
0.32 & 1.34259(10) & 1.8213(89) & 1.7423(15) & 1.8574(16) & 2.7142(25) \\
\hline
\end{tabular}}
\caption[a]{Masses for $\betapl=6.8$, $16^3\times 32$,
and $\betapl=7.1$, $14^3\times 28$ lattices.
Entry `-' means no good mass fits were possible.}
\la{table:mass1}
\end{table}

\begin{table}[t]
\centerline{
\begin{tabular}{|l|lllll|}
\multicolumn{6}{l}{Naik, $\betapl=7.4$, $16^3\times 32$} \\
\hline
$am_q$ &$\pi$     & $\pi_2$  & $\rho$    & $\rho_2$ & Nucleon   \\
\hline
0.02 & 0.37314(79) & 1.033(30) & 1.268(13) & 1.294(34) & 1.765(30)     \\
0.04 & 0.52238(70) & 1.0834(96) & 1.3139(89) & 1.383(14) & 1.920(15)   \\
0.08 & 0.72475(95) & 1.2517(75) & 1.4283(99) & 1.5035(71) & 2.1320(68) \\
0.16 & 1.0080(11) & 1.5591(97) & 1.6241(48) & 1.768(14) & 2.547(14)    \\
0.32 & 1.416801(79) & 2.146(23) & 1.9719(25) & 2.2396(79) & 3.1459(76) \\
\hline
\multicolumn{6}{l}{Kogut-Susskind, $\betapl=7.4$, $16^3\times 32$} \\
\hline
$am_q$ &$\pi$     & $\pi_2$  & $\rho$    & $\rho_2$ & Nucleon   \\
\hline
0.02 & 0.38080(74) & 0.9603(66) & 1.207(14) & 1.247(33) & 1.724(29)      \\
0.04 & 0.5280(12) & 1.0418(45) & 1.2449(63) & 1.337(43) & 1.855(13)	 \\
0.08 & 0.7297(10) & 1.1905(66) & 1.3270(28) & 1.4074(63) & 2.0282(53)	 \\
0.16 & 0.9994(11) & 1.4197(28) & 1.4774(25) & 1.5725(50) & 2.3174(56)	 \\
0.32 & 1.362520(74) & 1.7300(16) & 1.69637(81) & 1.7975(31) & 2.6656(35) \\
\hline
\multicolumn{6}{l}{} \\
\multicolumn{6}{l}{Naik, $\betapl=7.6$, $16^3\times 32$} \\
\hline
$am_q$ &$\pi$     & $\pi_2$  & $\rho$    & $\rho_2$ & Nucleon   \\
\hline
0.01 & 0.27294(23) & -          & 0.945(28) & -          & 1.36(15)    \\
0.02 & 0.38154(24) & 0.7513(89) & 1.0106(87) & 1.046(11) & 1.411(52)   \\
0.04 & 0.53125(25) & 0.8629(46) & 1.0911(56) & 1.1424(73) & 1.598(11)  \\
0.08 & 0.73613(23) & 1.0559(32) & 1.2420(53) & 1.312(11) & 1.852(40)   \\
0.16 & 1.01922(24) & 1.3667(35) & 1.4682(31) & 1.5441(56) & 2.256(15)  \\
0.32 & 1.42502(16) & 1.9038(93) & 1.8670(34) & 2.0072(69) & 2.9137(61) \\
\hline
\multicolumn{6}{l}{Kogut-Susskind, $\betapl=7.6$, $16^3\times 32$} \\
\hline
$am_q$ &$\pi$     & $\pi_2$  & $\rho$    & $\rho_2$ & Nucleon   \\
\hline
0.01 & 0.28090(23) & -          & 0.945(29) & -          & 1.28(14)     \\
0.02 & 0.39165(22) & 0.7508(90) & 0.9943(86) & 1.031(11) & 1.377(54)    \\
0.04 & 0.54237(23) & 0.8642(48) & 1.0621(48) & 1.190(15) & 1.577(10)    \\
0.08 & 0.74470(22) & 1.0462(31) & 1.2030(44) & 1.2705(91) & 1.804(33)   \\
0.16 & 1.01321(19) & 1.3092(27) & 1.3882(22) & 1.4474(38) & 2.187(20)   \\
0.32 & 1.36891(12) & 1.6677(32) & 1.6573(10) & 1.7287(19) & 2.5975(52)  \\
\hline
\multicolumn{6}{l}{} \\
\end{tabular}}
\caption[a]{Masses for $\betapl=7.4$, $14^3\times 28$, and
$\betapl=7.6$, $16^3\times 32$ lattices.}
\la{table:mass2}
\end{table}

\begin{table}[t]
\centerline{
\begin{tabular}{|l|lllll|}
\multicolumn{6}{l}{Naik, $\betapl=7.75$, $16^3\times 32$} \\
\hline
$am_q$ &$\pi$     & $\pi_2$  & $\rho$    & $\rho_2$ & Nucleon   \\
\hline
0.01 & 0.26943(34) & 0.5374(51) & 0.829(18)  & 0.862(12)  & 1.140(80) \\
0.02 & 0.37638(30) & 0.6175(42) & 0.8951(96) & 0.906(11)  & 1.265(26) \\
0.04 & 0.52339(20) & 0.7421(22) & 0.9701(47) & 1.015(13)  & 1.429(10) \\
0.08 & 0.72718(18) & 0.9428(14) & 1.1071(22) & 1.1479(77) & 1.6855(35)\\
0.16 & 1.01186(16) & 1.2587(17) & 1.3577(65) & 1.3963(26) & 2.1018(42)\\
\hline
\multicolumn{6}{l}{Kogut-Susskind, $\betapl=7.75$, $16^3\times 32$} \\
\hline
$am_q$ &$\pi$     & $\pi_2$  & $\rho$    & $\rho_2$ & Nucleon   \\
\hline
0.01 & 0.27794(33) & 0.5445(50) & 0.830(16)  & 0.863(11) & 1.174(11)   \\
0.02 & 0.38752(28) & 0.6182(22) & 0.8895(82) & 0.8898(53) & 1.282(10)  \\
0.04 & 0.53656(21) & 0.7504(21) & 0.9632(41) & 1.025(11) & 1.4225(90)  \\
0.08 & 0.73830(17) & 0.9460(13) & 1.0984(32) & 1.1351(54) & 1.6866(64) \\
0.16 & 1.00860(12) & 1.2320(16) & 1.3173(32) & 1.3550(41) & 2.0571(47) \\
\hline
\multicolumn{6}{l}{} \\
\multicolumn{6}{l}{Naik, $\betapl=7.9$, $16^3\times 32$} \\
\hline
$am_q$ &$\pi$     & $\pi_2$  & $\rho$    & $\rho_2$ & Nucleon   \\
\hline
0.005 & 0.18486(32) & 0.3806(67) & 0.6681(99) & 0.6831(75) & 0.920(12)\\
0.01 & 0.25907(28) & 0.4284(48) & 0.760(23) & 0.725(11) & 1.017(16)\\
0.02 & 0.36229(28) & 0.5054(37) & 0.7873(98) & 0.7678(53) & 1.0972(75)\\
0.04 & 0.50589(26) & 0.6400(20) & 0.8655(89) & 0.8546(49) & 1.2632(72)\\
0.08 & 0.70786(24) & 0.8506(16) & 1.0056(33) & 1.0104(28) & 1.5236(93)\\
0.16 & 0.99501(20) & 1.1710(14) & 1.2646(32) & 1.2790(17) & 1.9560(52)\\
\hline
\multicolumn{6}{l}{Kogut-Susskind, $\betapl=7.9$, $16^3\times 32$} \\
\hline
$am_q$ &$\pi$     & $\pi_2$  & $\rho$    & $\rho_2$ & Nucleon   \\
\hline
0.005 & 0.19160(32) & 0.3848(71) & 0.6704(96) & 0.6756(73) & 0.9114(67) \\
0.01 & 0.26809(28) & 0.4339(49) & 0.720(24) & 0.753(28) & 1.005(15)     \\
0.02 & 0.37383(31) & 0.5142(36) & 0.7767(87) & 0.796(11) & 1.0963(69)   \\
0.04 & 0.51985(29) & 0.6525(20) & 0.8599(44) & 0.8576(45) & 1.2703(65)  \\
0.08 & 0.72125(26) & 0.8624(15) & 1.0051(23) & 1.0115(26) & 1.5376(78)  \\
0.16 & 0.99577(21) & 1.1642(27) & 1.2401(19) & 1.2630(37) & 1.9388(54)  \\
\hline
\multicolumn{6}{l}{} \\
\end{tabular}}
\caption[a]{Masses for $\betapl=7.75$ and $\betapl=7.9$, $16^3\times 32$ 
lattices.}
\la{table:mass3}
\end{table}

The Naik hadron propagator calculation requires about 2 times more
CPU time than the Kogut-Susskind one.  The number of conjugate
gradient iterations is very similar for the Naik and the
Kogut-Susskind quarks, but since the Naik action \nr{naik} involves
about twice as many terms, the computational load is higher.  For
example, for the $\betapl=7.4$, $16^3\times 32$ lattices the number of
the conjugate gradient iterations for each source plane varies
approximately from 130 ($am_q = 0.32$) to 2050 ($am_q = 0.02$) for the
Kogut-Susskind and from 140 to 2400 for the Naik action, whereas the
CPU time per plane for $am_q=0.02$ is about 260 seconds for K-S and
600 seconds for Naik on the Intel Paragon using 32 nodes.

In order to find the best confidence levels of the propagator fits, we used
one-, two- and three-particle fitting functions, varying both the
beginning and the end of the fitting range.  All of the
fits use the full invariance matrix of the propagators.
We block together all of the propagators on each lattice, then,
in order to facilitate further analysis, we calculate the masses
using a single elimination jackknife procedure.  When fitting
each jackknife sample, we use the invariance matrix of the entire ensemble,
rather than recomputing the invariance matrix for each sample.


\begin{figure}[tb]
\centerline{\hspace*{-2cm}\epsfxsize=17cm\epsfbox{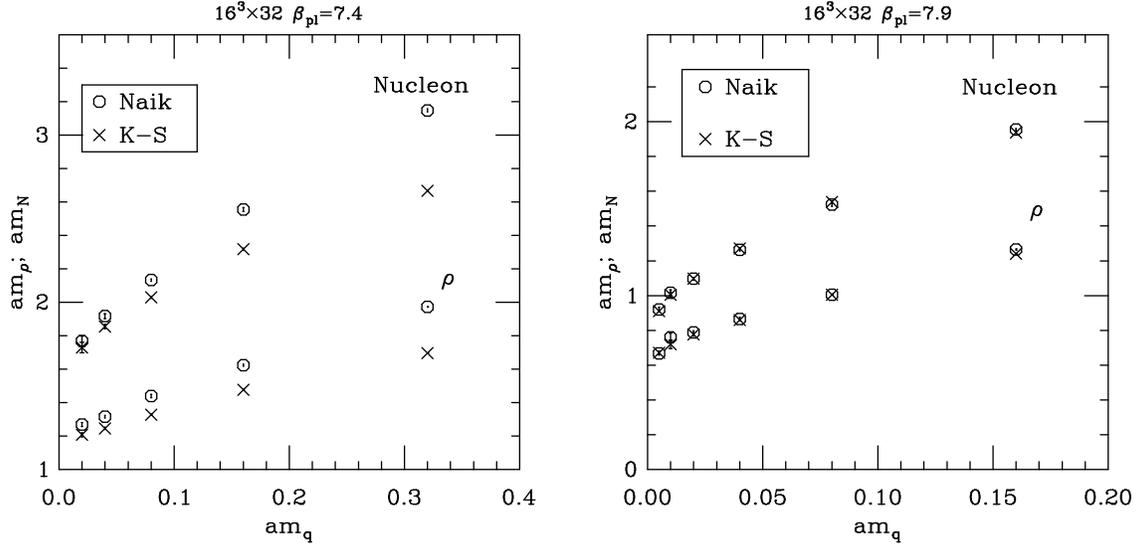}}
\vspace*{-16cm}
\caption[a]{Nucleon (upper) and $\rho$ (lower) masses as functions of
$am_q$ for $\betapl = 7.4$ and 7.9\@.} \la{fig:nucrho}
\end{figure}

\begin{figure}[tb]
\centerline{
\epsfxsize=10cm\epsfbox{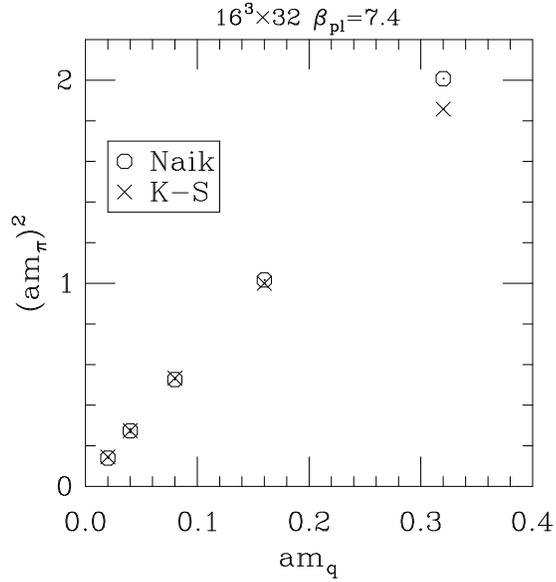}\hspace{2cm}}
\vspace*{-3cm}
\caption[a]{The pion mass squared as a function of $am_q$ for
$\betapl = 7.4$.} \la{fig:pisqr}
\end{figure}

A generic feature of the fits to the propagators is that one has to
use considerably larger minimum fit distance from the source with Naik
fermions than with the Kogut-Susskind fermions.  
The reason for this effect is probably the large extent in the 
imaginary time direction of the Naik
derivative operator in \eq\nr{naik} \cite{LuscherWeisz84}. 
The transfer matrix is well defined only
at imaginary time separations larger or equal to 3.
The ghost branch in the dispersion relation can
also cause short distance effects in the correlation function.

In \fig\ref{fig:nucrho} we summarise the nucleon and $\rho$ meson
masses from Tables \ref{table:mass2} and \ref{table:mass3} for
$\betapl=7.4$ and 7.9\@, and in \fig\ref{fig:pisqr} the pion mass
squared for $\betapl=7.4$.  The masses of the Naik hadrons in lattice
units tend to be larger than the Kogut-Susskind masses, but the
difference gets smaller with decreasing $am_q$ (approaching the chiral
limit) and increasing $\betapl$ (continuum limit).  At $\betapl=7.9$
the differences are barely discernible.  

\begin{figure}[tb]
\centerline{\hspace*{-2cm}\epsfxsize=17cm\epsfbox{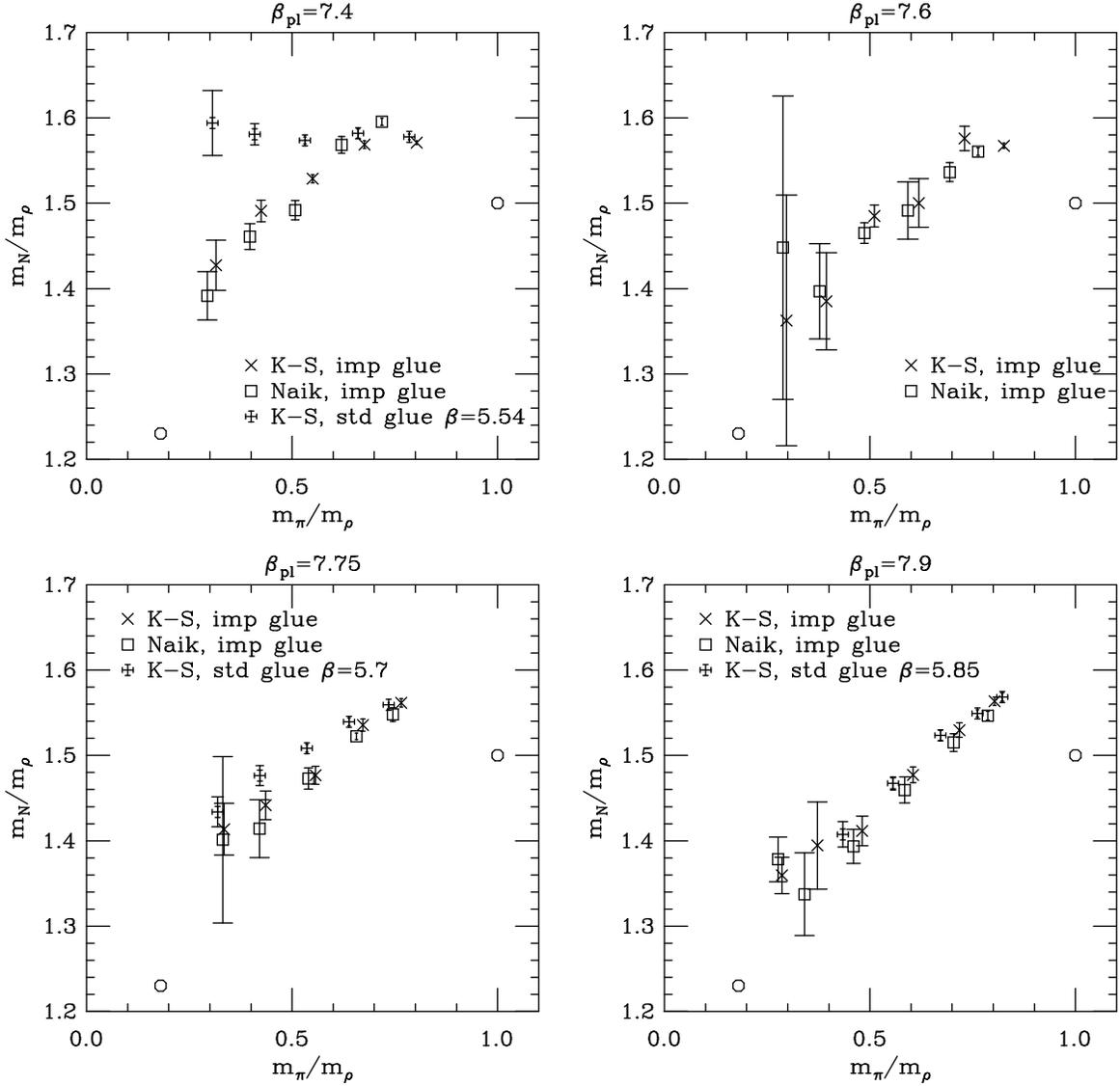}}
\vspace*{-8.3cm}
\caption[a] {The Edinburgh plots for $\betapl=7.4$, 7.6, 7.75 and
7.9\@.  The fancy crosses show the standard unimproved Kogut-Susskind
data at $\beta_{\rm Wilson} = 6/g^2=5.54$, 5.7 and 5.85
\cite{milcspectrum}; these correspond roughly to the same lattice
spacing (determined by $am_\rho$) as the improved $\betapl=7.4$, 7.75
and 7.9\@.  The small circles denote the physical limit ($m_\pi/m_\rho
\approx 0.18$) and the infinite quark mass limit ($m_\pi /
m_\rho=1$).}  \la{fig:edinburgh}
\end{figure}

Figure \ref{fig:edinburgh} shows the Edinburgh plots for
$\betapl=7.4$--7.9\@. 
The $m_N/m_\rho$ -ratios from
$\betapl=6.8$ and 7.1 exhibit typical strong coupling behavior:
the ratio $m_N/m_\rho$ remains roughly constant at around 1.5
when $m_q \rightarrow 0$ ($m_\pi/m_\rho \rightarrow 0$).
Only when $\betapl \ge 7.4$ do the data show an approach
to the vicinity of the physical value, and thus in the following we
concentrate on these couplings.  
For comparison, we also plot the $m_N/m_\rho$ mass ratio obtained with
standard non-improved Kogut-Susskind action at $\beta=5.54$, 5.7 and
5.85 \cite{milcspectrum}; these couplings correspond
roughly to the same lattice spacing as the improved action
at $\betapl=7.4$, 7.75 and 7.9\@.

It is interesting to note that despite the large differences in the masses
in lattice units in \fig\ref{fig:nucrho} (and Table
\ref{table:mass2}), in \fig\ref{fig:edinburgh} the $\betapl=7.4$ Naik
and Kogut-Susskind mass ratios lie practically on the same curve, with
the Naik values displaced slightly in the direction of smaller
$m_\pi/m_\rho$.  When the mass of the $\rho$ meson is used to set the
scale, from \figs\ref{fig:nucrho} and \ref{fig:edinburgh} we see that
for a given set of bare parameters the lattice spacing for the Naik
fermions is slightly larger than for the Kogut-Susskind
fermions, while the mass ratios are closer to the physical
values.  Conversely, if we want to investigate the same {\em physical
system\,} (mass ratio and physical volume), the Naik action enables us
to use a bit larger bare quark mass and smaller lattices (in lattice
units).  This effect becomes smaller as one gets closer to the
chiral limit $m_q\rightarrow 0$ and to the continuum limit
$\betapl\rightarrow\infty$; nevertheless, it is clearly observable
in all of the Edinburgh plots in \fig\ref{fig:edinburgh}.


\subsection{The chiral function and the continuum limit}

A convenient method to quantitatively measure the degree of improvement
in hadron spectroscopy is to study the lattice spacing dependence
(in units of $am_\rho$) of the ratio $m_N/m_\rho$ at some fixed
value of $m_\pi/m_\rho$.  This requires 
interpolation or extrapolation to the desired $m_\pi/m_\rho$ ratio.  We
perform this for each $\betapl$ separately with chiral fit functions.

The chiral fits are motivated by quenched chiral perturbation
theory ($Q\chi PT$), which gives $m_N$ and $m_\rho$ in a power series of
$m_\pi$ (+ logarithmic terms).  Since in the leading order $m_\pi^2
\propto m_q$, these become power series in $m_q^{1/2}$.  
For the standard Kogut-Susskind action, the chiral
extrapolations have been discussed in detail in \cite{milcspectrum}.

Extrapolation of the ratio $m_N/m_\rho$ to the chiral limit
$m_\pi\rightarrow 0$ or even to the physical limit $m_\pi/m_\rho
\approx 0.1753$ is sensitive to the form of the selected chiral
fit function ansatz.  The value of the ratio is much less sensitive in
the region $m_\pi/m_\rho \approx 0.4$--0.6, where the function
interpolates between measured mass values (see
\fig\ref{fig:edinburgh}).  Detailed comparison of the different
actions is feasible in this region.

We exclude the ``strong coupling'' runs at $\betapl=6.8$ and 7.1, and
fit $am_N(am_q)$ and $am_\rho(am_q)$ for $\betapl=7.4$--7.9 to the
chiral ansatz
\be
  am = c_0 + c_1\,am_q + c_{3/2} (am_q)^{3/2} + c_2 (am_q)^2 \,.
\la{chfit}
\ee
This function gives good fits at all 4 couplings (after excluding the
anomalous smallest $am_q$ value 0.005 from $\betapl=7.9$).  

        Q$\chi$PT for nucleons and vector mesons \cite{qxpt}
        implies the presence of an additional term $\propto
        am_\pi$, which corresponds to $(am_q)^{1/2}$ in the continuum.
        If one includes a term $\propto (am_q)^{1/2}$ in \nr{chfit},
        the fits invariably prefer a positive sign for the coefficient;
        whereas Q$\chi$PT gives a negative sign.  However, the appropriate
        pion mass in this term is actually the flavor singlet pion mass,
        which is not proportional to $(am_q)^{1/2}$ at fixed lattice
        spacing due to flavor symmetry breaking.  When this is taken
        into account, acceptable fits with a coefficient compatible
        both in sign and magnitude with Q$\chi$PT are possible.
        This is studied in detail for our standard gauge
        Kogut-Susskind hadrons in Ref.~\cite{milcspectrum}.
        Since such fits do not appear to change the 
        extrapolated/interpolated values significantly from \nr{chfit}, 
	but do increase the errors, we prefer to leave out the 
        $am_\pi$ term.

The error propagation is
taken into account by performing the (fully correlated) fits
separately to each of the jackknife blocks.

\begin{figure}[tb]
\centerline{\hspace*{-2cm}\epsfxsize=17cm\epsfbox{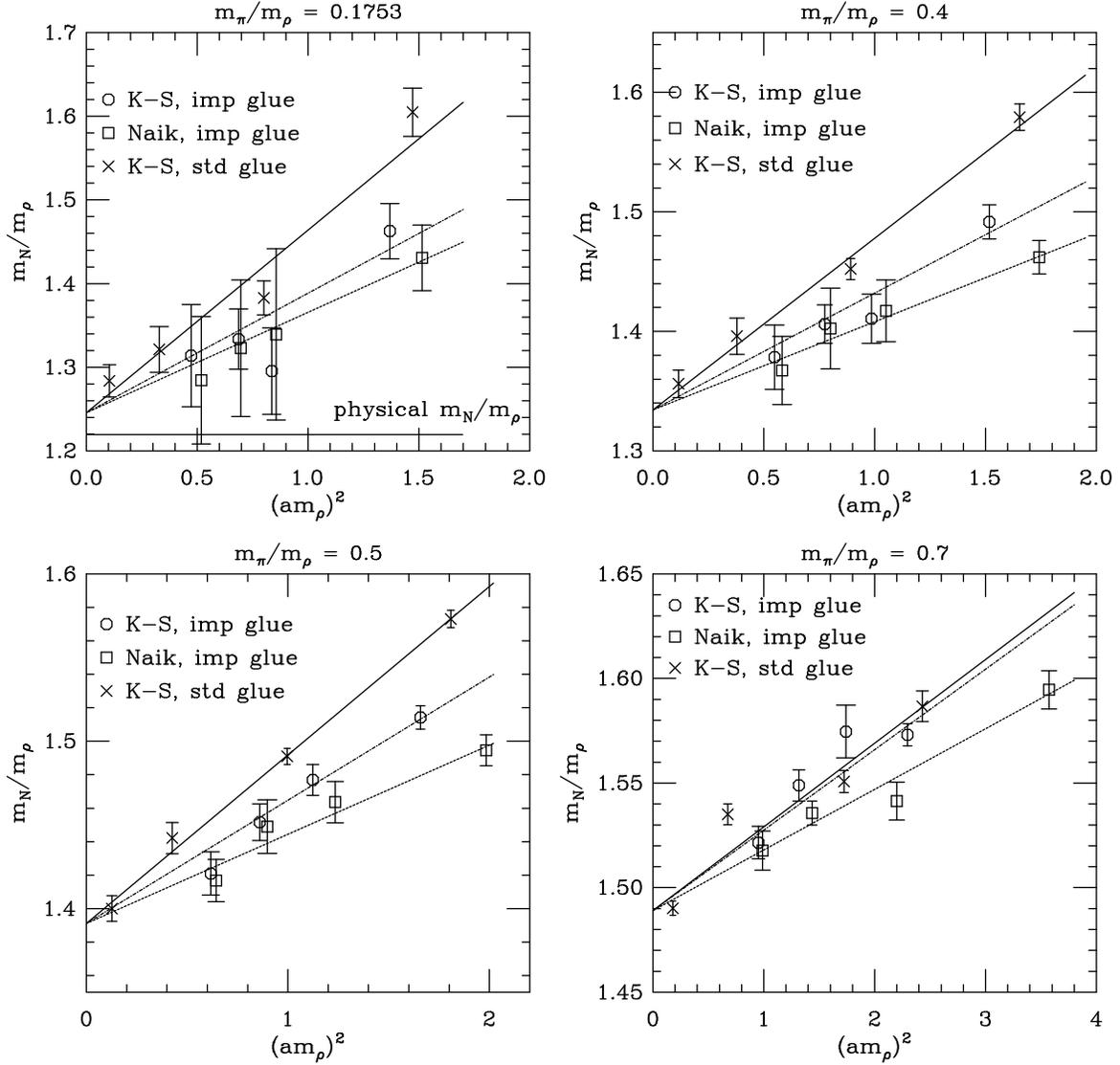}}
\vspace*{-8.3cm}
\caption[a]{The $m_N/m_\rho$ mass ratios as functions of the square of
the lattice spacing (in units of $(am_\rho)^2$), for $m_\pi/m_\rho =
0.1753$, 0.4, 0.5 and 0.7\@.  From left to right, the Naik and the
(improved gauge) Kogut-Susskind points correspond to $\betapl=7.9$,
7.75, 7.6 and 7.4; the standard Kogut-Susskind points to $\beta_{\rm
Wilson} = 6.15$, 5.85, 5.7 and 5.54\@.  The straight lines are linear
fits to the (from top to bottom) standard K-S, improved K-S and Naik
data, where the intercept at $a = 0$ in the last two fits is
fixed to the standard K-S value.}  \la{fig:chiral}
\end{figure}

In most cases, it would be possible to obtain acceptable fits also with
the simpler ansatz with either $c_{3/2}$ or $c_2$ fixed to zero.
However, while the full ansatz \nr{chfit} works quite well for the
standard Kogut-Susskind hadrons \cite{milcspectrum}, these simplified
functions do not.  In order to facilitate the comparisons between the
different actions, we retain the full chiral ansatz \nr{chfit} here.

The results of the chiral extrapolation/interpolation to $m_\pi/m_\rho
= 0.1753$ (physical), 0.4, 0.5 and 0.7 are shown in
\fig\ref{fig:chiral}; both for improved actions and for the standard
Kogut-Susskind action.  Since we expect the leading errors to be
$\cO(a^2)$, we plot the ratios against $(am_\rho)^2$.  Here
$am_\rho$ is calculated at the quark mass which yields the indicated value of
$m_\pi/m_\rho$.  We make a linear fit with respect to $(am_\rho)^2$ of the
standard Kogut-Susskind data, and, since in the continuum limit all of
the actions must yield equivalent results, we fit straight lines to
the improved Kogut-Susskind and Naik data, with the constraint that
the $a=0$ intercept is fixed to the standard Kogut-Susskind value.

We make the following observations:

\begin{itemize}
\item
In the intermediate $m_\pi/m_\rho = 0.4$ and 0.5 plots, the improved 
gauge nucleon to $\rho$ mass ratios are clearly closer to the continuum values than
the standard Kogut-Susskind ones.  Indeed, the $\betapl=7.9$ value
is very close to the standard Kogut-Susskind
$\beta_{\rm Wilson} = 6.15$ one, but with twice the lattice spacing
(albeit with larger statistical errors).  At large lattice
spacings ($\betapl=7.4$) the Naik fermions show smaller scaling
violation than
the improved gauge Kogut-Susskind fermions, but this difference
becomes very small when the lattice spacing is reduced.

\item
At the physical ratio $m_\pi/m_\rho = 0.1753$ the errors in
$m_N/m_\rho$ increase dramatically due to the extrapolation in 
$am_q$.  Nevertheless, we observe a pattern similar to that
at larger quark mass.

\item
When $m_\pi/m_\rho=0.7$ the quark mass $am_q$ becomes so large that
the chiral expansion \nr{chfit} does not converge well any more: the
highest power terms have the largest magnitude.  This leads to erratic
jumping of the points in the last panel of \fig\ref{fig:chiral}.
(note however the very small range of $m_N/m_\rho$ covered by this
plot).

\item
As the quark mass is lowered, the difference between the two types of
quarks in the improved gluonic fields is reduced.  Thus, at 
$m_\pi/m_\rho=0.7$ most of the improvement of $m_N/m_\rho$ is attributable 
to the Naik improvement, whereas near the physical quark mass, most of the
improvement comes from the gluonic action.
A large part of the Naik improvement is due to the larger ($am_\rho$) and
hence a larger lattice spacing.  If one uses the string tension to set
the scale the difference between the Naik and the Kogut-Susskind actions
becomes smaller.

\item 
The linearity (against $(am_\rho)^2$) of the standard Kogut-Susskind
$m_N/m_\rho$-ratio clearly supports the notion that the scaling violations
behave as $\cO(a^2)$.  When $m_\pi/m_\rho \le 0.5$, the
constrained linear fits to the improved gauge Kogut-Susskind and Naik
data have confidence levels better than 0.5, certainly quite
compatible with $\cO(a^2)$ leading scaling violations.  The
magnitude of the violations -- the slope of the line -- for the Naik
data is only about $1/2$ of the standard Kogut-Susskind value, whereas
the improved gauge Kogut-Susskind has a slightly larger slope than
Naik.

\item
We can also test whether the data would allow for $\cO(a^3)$
scaling violations for the Naik action.  When $m_\pi/m_\rho =0.5$ a
constrained fit of form $A + B(am_\rho)^3$ (where again $A$ is set to
the $a=0$ intercept of the standard Kogut-Susskind data) does {\em
not} fit the Naik data well: the confidence level is only 0.15 (as
opposed to 0.75 before).  This disfavors the leading $\cO(a^3)$
errors.  For smaller $m_\pi/m_\rho$-ratios the statistical errors
become larger and this analysis cannot distinguish the fits.

\item
To check consistency, we can relax the constraint at $a=0$ and fit
independent straight lines to all datasets.  When $m_\pi/m_\rho \le
0.5$ the intercepts at $a=0$ are compatible for all cases, \ie,
within 1 standard deviation of each other.

\end{itemize}

The lattice spacing and the size of the system in physical units can
be obtained by extrapolating $am_\rho$ to the physical $m_\pi/m_\rho$-ratio
and setting $m_\rho=770$\,MeV\@.  These are given in Table
\ref{tab:size}.

\begin{table}[ht]
\centerline{
\begin{tabular}{l|llllll}
\hline
 $\betapl$ & 6.8  & 7.1  & 7.4  & 7.6  & 7.75  & 7.9 \\
\hline
 $a$ (fm)    & 0.37 & 0.35 & 0.31 & 0.24 & 0.21  & 0.19 \\
 Size (fm)   & 5.9  & 5.0  & 5.0  & 3.8  & 3.4   & 3.0 \\
\hline
\end{tabular}}
\caption[0]{The lattice spacing and the box size in physical
units.}\la{tab:size}
\end{table}

The numbers in Table \ref{tab:size} have been calculated with the
Naik quark action; for the Kogut-Susskind action the lattice spacings and
the box sizes would be fractionally smaller.  The box sizes are considerably
larger than 2\,fm (with the possible exception of $\betapl=7.9$),
so that we can safely ignore the finite size effects \cite{Steven}.
For the weakest coupling and the smallest quark mass, the product
$m_\pi \times (\mbox{Lattice size})$ is approximately 3.0.

\begin{figure}[tb]
\centerline{
\hspace*{-2cm}\epsfxsize=17cm\epsfbox{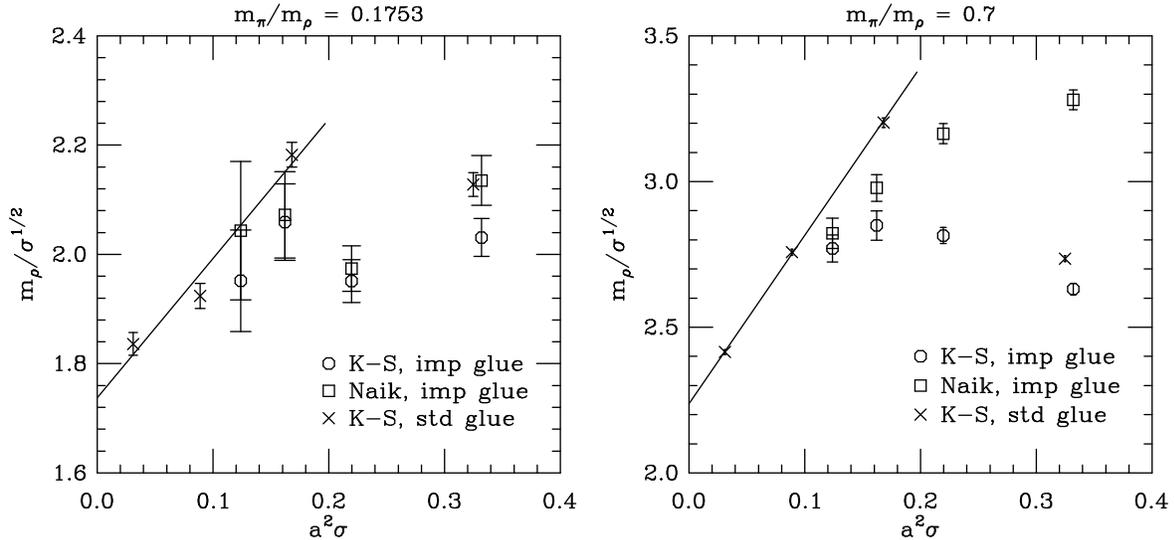}}
\vspace*{-16cm}
\caption[a]{The ratio $m_\rho/\sqrt{\sigma}$ against $a^2\sigma$
at $m_\pi/m_\rho = 0.1753$ and 0.7.} \la{fig:rhosigma}
\end{figure}

Besides the mass of the $\rho$-meson, the square root of the string
tension is commonly used to determine the lattice spacing.
In \fig\ref{fig:rhosigma} the ratio $m_\rho/\sqrt{\sigma}$ is plotted
against $a^2\sigma$.  Here $am_\rho$ is evaluated at the physical
$m_\pi/m_\rho=0.1753$ and at 0.7\@.  The string tension for the standard
gauge action is interpolated from the data in the literature \cite{tension};
for the improved gauge action \nr{gaugeaction} it has been
measured by the SCRI group \cite{scri}.

Since the scale violations are expected to behave as $\cO(a^2)$,
the $m_\rho/\sqrt{\sigma}$ ratio should behave linearly as a function
of $a^2\sigma$.  Indeed, the standard Kogut-Susskind data shows good
linearity up to $a^2\sigma = 0.17$ ($\beta_{\rm Wilson} = 5.7$).
However, at stronger coupling (5.54) the ratio falls strongly off the
linear behavior.  We fit a straight line to the three weakest coupling
datapoints, the intercepts at $a=0$ are 1.738(25) at the physical
$m_\pi/m_\rho=0.1753$ and 2.238(11) at 0.7\@ (the errors quoted here
are only statistical).  These results are consistent with the SCRI
group (preliminary) Wilson and clover fermion mass ratios \cite{scri}.

The improved gauge Kogut-Susskind and Naik data still seem to reside
completely in the ``strong coupling region'', although there is some
indication that at weaker couplings the ratios would bend to the
direction of the line defined by the standard Kogut-Susskind data.
Extrapolation of the improved action ratios to the continuum limit
is clearly not justified.  

The non-linearity in $m_\rho/\sqrt{\sigma}$ is somewhat surprising,
when we compare it against the purely hadronic observables in
\fig\ref{fig:chiral}.  This lends support to the view that a large
part of the scaling violations cancel in the hadronic ratios, and
justifies the use of $am_\rho$ as the scale factor in purely hadronic
observables.

\subsection{Lorentz symmetry} 

As discussed in Sec.~\ref{sec:properties}, the free quark continuum 
dispersion relation is approximated much better by the Naik action
than by the standard Kogut-Susskind action.  At very high
temperatures, deep in the quark-gluon plasma phase, the quarks
behave approximately as free particles, and the Naik
action improves QCD thermodynamics \cite{Karsch96}.
However, {\it a priori} it is not clear whether the dispersion
relation of hadronic states is improved.

\begin{table}[t]
\centerline{
\begin{tabular}{|c|ll|ll|}
\hline
&\multicolumn{2}{|c|}{Kogut-Susskind $\pi$} 
&\multicolumn{2}{|c|}{Naik $\pi$} \\
\hline
${\bf k}L/(2\pi)$& 
$aE({\bf k})$& $c^2({\bf k})$ & $aE({\bf k})$& $c^2({\bf k})$ \\
\hline
(0,0,0)& 0.53521(17) &  -        & 0.52625(15) & - \\
(0,0,1)& 0.65223(56) & 0.9010(45)& 0.65111(50) & 0.9532(39)\\
(0,1,1)& 0.74710(63) & 0.8809(30)& 0.75474(76) & 0.9489(36)\\
(1,1,1)& 0.82655(88) & 0.8575(31)& 0.8411(11)  & 0.9305(40)\\
(0,0,2)& 0.8817(14)  & 0.7959(40)& 0.9087(26)  & 0.8897(77)\\
(0,2,2)& 1.0925(32)  & 0.7353(57)& 1.1549(44)  & 0.8566(83)\\
(2,2,2)& 1.275(11)   & 0.723(15) & 1.395(19)   & 0.902(29)\\
\hline
\multicolumn{5}{c}{}\\
\hline
&\multicolumn{2}{|c|}{Kogut-Susskind $\rho$}
&\multicolumn{2}{|c|}{Naik $\rho$} \\
\hline
${\bf k}L/(2\pi)$& 
$aE({\bf k})$& $c^2({\bf k})$ & $aE({\bf k})$& $c^2({\bf k})$ \\
\hline
(0,0,0)& 1.2411(69) & -       &  1.3065(79)& -       \\
(0,0,1)& 1.262(27)  & 0.34(44)&  1.3489(80)& 0.73(15)\\
(0,1,0)& 1.289(13)  & 0.79(21)&  1.362(16) & 0.97(28)\\
(0,1,1)& 1.298(13)  & 0.47(11)&  1.385(19) & 0.68(16)\\
(1,1,0)& 1.320(18)  & 0.65(16)&  1.385(12) & 0.68(12)\\
\hline
\end{tabular}}
\caption[a]{The energy of the $\pi$ and $\rho$ meson states at finite
momentum ${\bf k} = {\bf n} 2\pi/L$, and the `speed of light squared'
$c^2({\bf k}) = (E^2({\bf k}) - E^2(0))/{\bf k}^2$, for $\betapl=7.4$,
$am_q = 0.04$, $16^3\times 32$ lattice.}  \la{table:csqr}
\end{table}

\begin{figure}[tb]
\centerline{
\hspace*{-2cm}\epsfxsize=11cm\epsffile{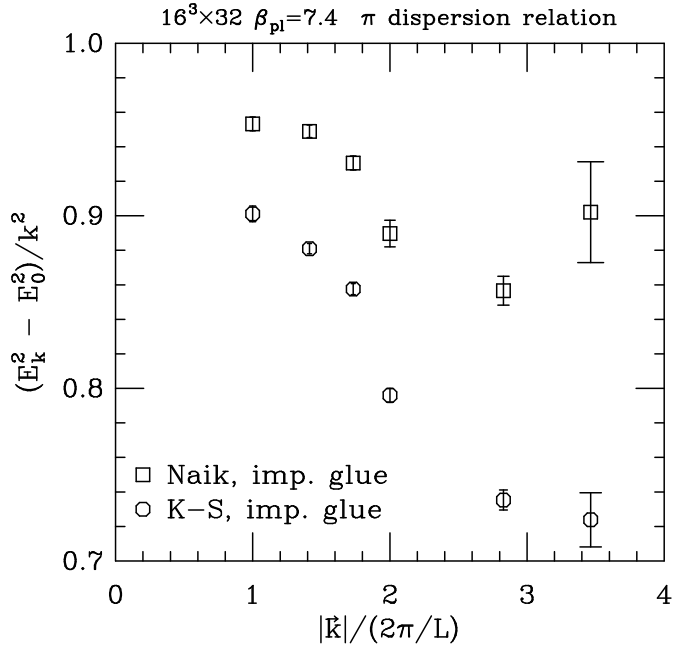}}
\vspace*{-1cm}
\caption[a]{The `speed of light squared', calculated
from the pion dispersion relation, for Naik and K-S pions.}
\la{fig:csqr}
\end{figure}

We test hadron dispersion relations by measuring the
energy of the $\pi$- and $\rho$-meson states with finite
spatial momenta on 100 lattices with $\betapl=7.4$, $am_q=0.04$ and
volume $16^3\times 32$.  We use 4 (finite momentum) wall 
sources, separated by 8 lattice units.

The source operators are constructed as follows: first, we take a zero
momentum wall source, which is 1 for a particular source color at each
spatial slice at the source time.  This is used as a source for the
conjugate gradient to compute the quark propagators.  Then this wall
source is multiplied by the momentum dependent phase factor $\exp( i
{\bf k} \cdot {\bf x} )$, by the sign factors (depending on the
location in the $2^4$ flavor hypercube) to select the desired meson,
and by an extra $(-1)^{\sum_\mu x_\mu }$ corresponding to $\gamma_5$.
This is used as a source for the conjugate gradient to compute the
antiquark propagator.  The sink operator is similar, except that the
quark and antiquark propagators are multiplied together with the
appropriate phase and sign factors before summing over spatial points,
corresponding to a local sink.

For pions, we use momentum vectors pointing to 3 different directions:
${\bf k}L/(2\pi) = (0,0,1)$, (0,1,1), (1,1,1), and these multiplied by
2.  For the $\rho$-meson, we use (the lattice analog of) the vector
operator $\bar\psi\gamma_3\psi$, and we expect that the dispersion
relation may be different along the $z$-axis direction and
perpendicular to it.  Therefore, for $\rho$ we use ${\bf k}L/(2\pi) =
(0,0,1)$, (0,1,0), (0,1,1) and (1,1,0).  The signals for higher
momenta are too noisy to be useful.  The results are listed in Table
\ref{table:csqr}.

The violation of Lorentz invariance can be quantified by
measuring the `speed of light' with the continuum dispersion relation
\be
    c^2({\bf k}) = \frac{E^2({\bf k}) - E^2(0)}{{\bf k}^2}\,.
\ee
The deviation of $c^2$ from unity directly measures the violation of
Lorentz invariance.  The results are shown in Table
\ref{table:csqr} and in \fig\ref{fig:csqr} (for pions).  The
Naik pions show a clear improvement of $c^2$ over the Kogut-Susskind
pions: the deviation from unity is reduced approximately by half.  The
results for the $\rho$-mesons seem to indicate a dependence on the
direction of the momentum (parallel or perpendicular to $z$).  Also,
here the $c^2$ is closer to unity for the Naik mesons; however, the
statistical errors are so large that we cannot make definite
statements about the improvement.

\subsection{Flavor symmetry}

\begin{figure}[tb]
\centerline{
\hspace*{-2cm}\epsfxsize=10cm\epsffile{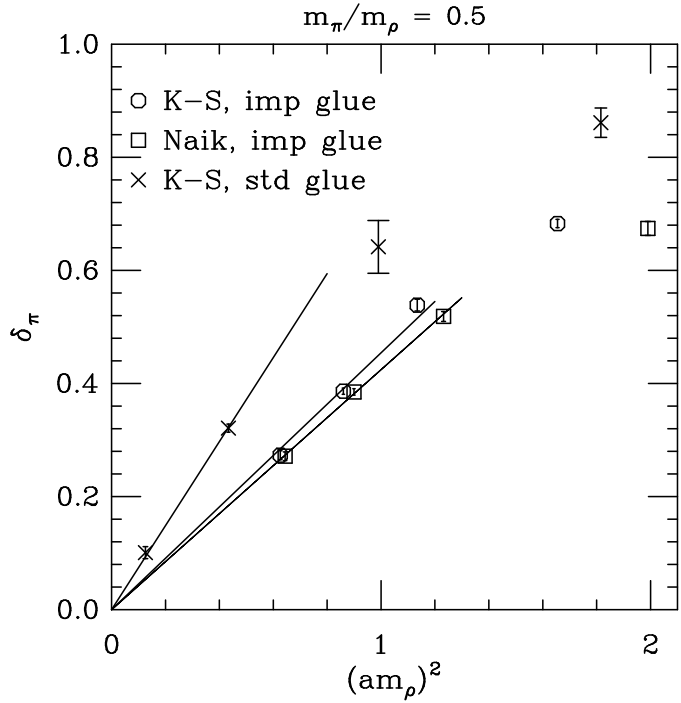}}
\vspace*{-1cm}
\caption[a]{Flavor symmetry breaking parameter
$\delta_\pi=(m^2_{\pi_2}-m^2_\pi)/(m^2_\rho-m^2_\pi)$,
interpolated to $m_\pi/m_\rho = 0.5$, as a function
of $(am_\rho)^2$.  The data correspond to the same
values of $\beta$ as in \fig\ref{fig:chiral}.   
The straight lines are linear fits 
to 2 (standard gauge) or 3 (improved gauge) points with the smallest
lattice spacings, constrained to go through the origin.}\la{fig:delta}
\end{figure}

The restoration of flavor symmetry can be discerned by
investigating the mass differences between $\pi$ and $\pi_2$ mesons.
The first particle is the Goldstone boson corresponding to the
spontaneously broken U(1)$\times$U(1) chiral symmetry and it becomes
massless when $am_q\rightarrow 0$ even at a finite lattice spacing
(\fig\ref{fig:pisqr}).  In comparison, the $\pi_2$ mesons remain
massive in the chiral limit, and become massless only when both
the chiral and the continuum limits are taken.

We use the dimensionless quantity
\be
  \delta_\pi = \fr{m^2_{\pi 2} -
m^2_\pi}{m^2_\rho - m^2_\pi}
\ee
to measure flavor symmetry breaking.  For the standard Kogut-Susskind
quark action, this quantity is almost independent of the bare quark
mass $am_q$ at small lattice spacings.  In \fig\ref{fig:delta}, we show
$\delta_\pi$ for $\betapl\ge 7.4$ improved gauge Naik and
Kogut-Susskind hadrons, together with the unimproved glue
Kogut-Susskind values, as functions of $(am_\rho)^2$.  The data is
interpolated to $m_\pi/m_\rho=0.5$ (compare to the third panel in
\fig\ref{fig:chiral}).  With this constraint, the flavor symmetry
breaking parameter reduces to $\delta_\pi = (m^2_{\pi2}/m^2_\pi -
1)/3$.

At this value of $m_\pi/m_\rho$, we observe that the flavor symmetry
violation at small $a$ is reduced by $\approx 45$\% due to the
improved gauge action.  When the Naik fermions are used, $\delta_\pi$
is slightly smaller than with the Kogut-Susskind fermions.
However, this situation would become reversed, if we used $a\sqrt{\sigma}$
instead of $am_\rho$ to set the scale.

Figure \ref{fig:delta} clearly indicates that the leading flavor symmetry 
breaking terms are proportional to $a^2$ for all of the actions studied.
The region linear in $(am_\rho)^2$ extends to larger lattice
spacings with the improved gauge.

A successful additional improvement of the Kogut-Susskind flavor symmetry
is the MILC ``fat link'' fermion action \cite{fatlinks}.  That action
substitutes the standard gauge links with smeared average links in
the fermion hopping terms.  The averaging process improves the flavor
symmetry dramatically, the improvement being roughly comparable both
for the standard Kogut-Susskind and the Naik action.  These
observations indicate the importance of the coupling of fermions
to the gauge fields for the flavor symmetry of the staggered action.
The Naik action \nr{naik} can be interpreted naively as a
straightforward improvement of the (free) fermion dispersion relation.

  The improvement in flavor symmetry from the Symanzik improved gauge
  action, like the improvement from the fat link quark action, can be
  understood as a suppression of the effects of high momentum gluons.
  Gluons with momentum near $\pi/a$ scatter quarks from one corner of the
  Brillouin zone to another, which is roughly equivalent to changing their
  flavor \cite{Lepage,LagaeSinclair}.  
The suppression of the high momentum gluons becomes evident
when the gauge action is expanded to quadratic
order in the vector potential $A_\mu$
(where the lattice variable $U_\mu(x) = \exp[-igaA_\mu(x)]\,$).
Using the shorthand notations
\be
  \hat k_\mu = 2\sin {\fr12} ak_\mu, 
  \h \hat k^2 = \sum_\mu \hat k_\mu^2\,,
\ee
and
\be
  f_{\mu,\nu}(k) = \hat k_\mu A_\nu(k) - \hat k_\nu A_\mu(k)\,,
\ee
the quadratic part of the action \nr{gaugeaction} can be written
in the form \cite{Weisz}
\be
S^{(2)} = \frac{1}{2} a^2 \sum_{k;\mu<\nu} 
   f_{\mu,\nu}(k) f_{\mu,\nu}(-k) 
     \left[ \cpl + 8 \crt + 16 \cpg 
       - (\crt - \cpg) (\hat k_\mu^2 + \hat k_\nu^2)
       - \cpg \hat k^2\right]\,.
\la{a2action}
\ee
Here the coefficients $c_i$ denote the relative strength
of the three terms in the action.  They are related
to coefficients $\beta_i$ through $c_i\,6/g^2 = \beta_i$.
As an overall normalization we require that the constant term within
the brackets equals to one: $\cpl+8\crt+16\cpg=1$.

For the Wilson gauge action the coefficients are $\cpl=1$,
$\crt=\cpg=0$, whereas for the improved action $\cpl > 1$ and
$\crt,\cpg < 0$.  With the improved gauge, the non-constant
terms within the brackets in \eq\nr{a2action} increase the
action for modes close to the edges of the Brillouin zone ($k_\mu
\approx \pm \pi/a$, for at least one $\mu$).

As a simple example we consider a momentum vector parallel to one of
the lattice axes and at the edge of the zone.  In this case the term
within the brackets in \eq\nr{a2action} reduces to $[1 - 4 \crt]$.
At $\betapl=7.4$, using \eqs(\ref{betart},\ref{betapg}),
Table~\ref{tab:runs} and the normalization condition above, the
coefficent $\crt$ has a value $\approx -0.26$.  When compared to the
Wilson gauge action ($\crt=0$), this more than doubles the action
difference of the modes close to the edge of the zone and near the
origin $k=0$.  When the lattice spacing is reduced the suppression of
the modes near the edge of the zone increases rapidly, while the
relative difference between the actions becomes smaller.  At tree
level, the coefficients assume values $\cpl=5/3$, $\crt=-1/12$ and
$\cpg=0$, still yielding a 33\% difference of the action at $k_\mu =
\pi/a$.

\section{Conclusions}

We investigate improvement of the quenched
light hadron mass spectrum using a tadpole-improved
staggered Naik action \nr{naik}, which at the tree level does not 
have $\cO(a^2)$ errors.
Correspondingly, we use $\cO(a^2)$ tadpole improved gauge action to
generate the gauge configurations.  Using the same gauge action for
both the Kogut-Susskind and the Naik calculations allows us to separate
the effect of the fermionic improvement from the improvement of the
gauge action.  The latter is studied by comparing the results
presented here to our standard Kogut-Susskind results
\cite{milcspectrum}.

We find that improvement of the gauge action has a
significant effect on the hadron spectrum: when $m_\pi/m_\rho \sim
0.5$, the nucleon to $\rho$-meson mass ratio is roughly 50\,\% closer
to the continuum value with the improved gauge than with the standard
gauge action.
 Thus, the scale violations with the standard gauge spectroscopy
 are at roughly the same level as with the improved gauge at
 about $1.4$ times the lattice spacing.
Using the improved gauge action, the Naik quark 
action has
smaller scaling violations than the Kogut-Susskind action, although
the difference becomes small when the quark mass is reduced.
Similarly, improving the gauge action reduces the amount of
flavor symmetry breaking, but using the Naik action yields
little further gains.
For both of the actions the flavor symmetry can be further improved with the
`fat link' procedure \cite{fatlinks}.

The biggest improvement provided by the Naik action comes from the
improved Lorentz invariance of the hadronic states.  This is best evidenced
by the $\pi$-meson dispersion relation, which is much closer to
the continuum behavior when the Naik action is used.  This property
may be especially significant for nonzero temperature simulations, where
the hadronic and/or quark degrees of freedom typically have large momenta.
Thus, when one strives for higher precision in staggered quark
simulations, an economical solution can be found from the combination 
of an improved Naik-like quark action together with the fat links.

\subsubsection*{Acknowledgements}

Discussions with E. Laermann are gratefully acknowledged.
This work was supported by the U.S. Department of Energy under grants
DE--AC02--76CH--0016,
DE--AC02--86ER--40253,
DE--FG02--91ER--40661, 
DE--FG02--91ER--40628, 
DE--FG03--95ER--40894, 
DE--FG03--95ER--40906, 
DE--FG05--85ER250000,  
DE--FG05--96ER--40979, 
and National Science Foundation grants
NSF--PHY96--01227 and     
NSF--PHY97--22022.        
The computations have been performed on the
Intel Paragon at Indiana University, on the DEC Alpha server and Cray T3E
at the Pittsburgh Supercomputing Center, on the Intel Paragon and
Cray T3E at the San Diego Supercomputing Center, and on the CM-2 at SCRI\@.


\begin{thebibliography}{99}

\bibitem{Symanzik83}
K. Symanzik, in ``Recent Developments in Gauge Theories'', eds.
G. 't Hooft \etal 313 (Plenum, New York, 1980);
Nucl. Phys. {\bf B226} 187 (1983).

\bibitem{LuscherWeisz}
M. L\"uscher and P. Weisz, Comm. Math. Phys. {\bf 97} 19 (1985);
Phys. Lett. {\bf 158B} 250 (1985).

\bibitem{LepageMackenzie}
G. P. Lepage and P. B. Mackenzie, Phys. Rev. {\bf D48} 2250 (1993)
[{\tt hep-lat/9209022}].

\bibitem{DeGrand}
T. DeGrand, A. Hasenfratz, P. Hasenfratz and F. Niedermayer,
Nucl. Phys. {\bf B454} 587 (1995) [{\tt hep-lat/9506030}];
Nucl. Phys. {\bf B454} 615 (1995) [{\tt hep-lat/9506031}].

\bibitem{perfectKS}
W. Bietenholz and U.-J. Wiese, Nucl. Phys. {\bf B} (Proc. Suppl.) {\bf 34}
(1994) 516 [{\tt hep-lat/9311016}];
H. Dilger, Nucl. Phys. {\bf B490} (1997) 331 [{\tt hep-lat/9610029}];
W. Bietenholz, R. Brower, S. Chandrasekharan and U.-J. Wiese,
Nucl. Phys. {\bf B495} (1997) 285 [{\tt hep-lat/9612007}].

\bibitem{Niedermayer96}
F.~Niedermayer, review in the proceedings of ``Lattice '96'', 
Nucl. Phys. {\bf B} (1997) {\bf 53} 56 (1997) [{\tt hep-lat/9608097}].

\bibitem{Naik}
S. Naik, Nucl. Phys. {\bf B316} 238 (1989).

\bibitem{milc_naik}
C. Bernard \etal (MILC collaboration),
Nucl. Phys. {\bf B} (Proc. Suppl.) {\bf 53} 212 (1997) [{\tt hep-lat/9608102}];

to appear in the proceedings of ``Lattice '97'',
Nucl. Phys. {\bf B} (Proc. Suppl.) 
[{\tt hep-lat/9711013}].

\bibitem{fatlinks}
T. Blum, C. DeTar, S. Gottlieb, K. Rummukainen,
U.M. Heller, J. Hetrick, D. Toussaint, R.L. Sugar, M. Wingate
Phys. Rev. {\bf D55} 1133 (1997) [{\tt hep-lat/9609036}].

\bibitem{Karsch96}
F. Karsch, B. Beinlich, J. Engels, R. Joswig, E. Laermann,
A. Peikert and B. Petersson,
Nucl. Phys. {\bf B} (Proc. Suppl.) {\bf 53} 413 (1997) [{\tt hep-lat/9608047}].

\bibitem{milcspectrum}
C. Bernard \etal (MILC collaboration),
to appear in the proceedings of the workshop 
``Lattice QCD on Parallel Computers'', (Tsukuba 1997),
IUHET-366 [{\tt hep-lat/9707014}];
full report in
{\em Continuum Limit of Lattice QCD with Staggered Quarks in the Valence 
Approximation --- A Critical Role for the Chiral Extrapolation},
in preparation.

\bibitem{Alford95}
M. Alford, W. Dimm, G.P. Lepage, G. Hockney and P.B. Mackenzie,
Phys. Lett. {\bf B361} 87 (1995)
[{\tt hep-lat/9507010}].

\bibitem{KSmomentum}
H.S. Sharatchandra, H.J. Thun, P. Weisz 
Nucl. Phys. {\bf B192} 205 (1981);
M.F.L. Golterman and J. Smit, Nucl. Phys. {\bf B245} 64 (1984).

\bibitem{KlubergStern}
H. Kluberg-Stern, A. Morel, O. Napoly, B. Petersson,
Nucl. Phys. {\bf B220} [FS8] 447 (1983).

\bibitem{Sharpe}
S. Sharpe, Nucl. Phys. {\bf B} (Proc. Suppl.) 34 403 (1994) 
[{\tt hep-lat/9312067}].

\bibitem{Luo96}
Y. Luo, Phys. Rev. {\bf D 55} 353 (1997) [{\tt hep-lat/9604025}].

\bibitem{Luo97}
Y. Luo, Columbia University preprint CU-TP-186 [{\tt hep-lat/9702013}].

\bibitem{p4}
F. Karsch, 
to appear in the proceedings of the workshop 
``Lattice QCD on Parallel Computers'', (Tsukuba 1997) [{\tt hep-lat/9706006}];
A. Peikert, B. Beinlich, A. Bicker, F. Karsch and E. Laermann,
to appear in Nucl. Phys. B(Proc. Suppl.) [{\tt hep-lat/9709157}].

\bibitem{LuscherWeisz84}
M. L\"uscher and P. Weisz, Nucl. Phys. {\bf B240} 349 (1984). 

\bibitem{qxpt}   
J. Labrenz, S. Sharpe, Phys. Rev. {\bf D54} 4595 (1996) [{\tt hep-lat/9605034}];
M. Booth, G. Chiladze and A. Falk, Phys. Rev. {\bf D55} 3092 (1997) 
[{\tt hep-ph/9610532}].

\bibitem{Steven}
S. Gottlieb, Nucl. Phys. {\bf B} (Proc. Suppl.) {\bf 53} 155 (1997) 
[{\tt hep-lat/9608107}].

\bibitem{tension}
G. Bali and K. Schilling,
Phys. Rev. {\bf D46} 2636 (1992);
G. Boyd, J. Engels, F. Karsch, E. Laermann, C. Legeland,
M. L\"utgemeier and B. Petersson,
Nucl. Phys. {\bf B469} 419 (1996) [{\tt hep-lat/9602007}];
R.G. Edwards, U.M. Heller and T.R. Klassen, 
Florida State University preprint 
FSU-SCRI-97-122 [{\tt hep-lat/9711033}]

\bibitem{scri}
S. Collins, R.G. Edwards, U.M. Heller and J. Sloan,
in proceedings of the International Conference ``Multiscale Phenomena
and Their Simulation'', Bielefeld, Germany 1996, 
eds. F. Karsch, B. Monien and H. Satz,
World Scientific (1997) [{\tt hep-lat/9611022}].

\bibitem{Lepage}
P. Lepage, 
to appear in proceedings of the workshop
``Lattice QCD on Parallel computers'' (Tsukuba, 1997) 
[{\tt hep-lat/9707026}].

\bibitem{LagaeSinclair}
J.F. Laga\"e and D.K. Sinclair, 
to appear in the proceedings of ``Lattice '97'', 
Nucl. Phys. {\bf B} (Proc. Suppl.) 
[{\tt hep-lat/9709035}].

\bibitem{Weisz}
P. Weisz, Nucl. Phys. {\bf B212} 1 (1983).


\end{thebibliography}
\end{document}